# Boosting room temperature magneto-ionics in $Co_3O_4$

Julius de Rojas[1], Alberto Quintana[2], Aitor Lopeandía[1], Joaquín Salguero[3], José L. Costa-Krämer[3], Llibertat Abad[4], Maciej O. Liedke[5], Maik Butterling[5], Andreas Wagner[5], Lowie Henderick[6], Jolien Dendooven[6], Christophe Detavernier[6], Jordi Sort[1,7]* & Enric Menéndez[1]*

[1]Departament de Física, Universitat Autònoma de Barcelona, E-08193 Cerdanyola del Vallès, Spain

[2]Department of Physics, Georgetown University, Washington, D.C. 20057, United States

[3]IMN-Instituto de Micro y Nanotecnología (CNM-CSIC), Isaac Newton 8, PTM, 28760 Tres Cantos, Madrid, Spain

[4]Institut de Microelectrònica de Barcelona, IMB-CNM (CSIC), Campus UAB, E-08193 Bellaterra, Spain

[5]Institute of Radiation Physics, Helmholtz-Center Dresden-Rossendorf, Dresden 01328, Germany

[6]Department of Solid State Sciences, CoCooN, Ghent University, Krijgslaan 281/S1, 9000 Ghent, Belgium

[7]Institució Catalana de Recerca i Estudis Avançats (ICREA), Pg. Lluís Companys 23, E-08010 Barcelona, Spain

*email: jordi.sort@uab.cat (J. Sort), enric.menendez@uab.cat (E. Menéndez)

**Abstract**

Voltage control of magnetism through electric field-induced oxygen motion (magneto-ionics) could represent a significant breakthrough in the pursuit for new strategies to enhance energy efficiency in a large variety of magnetic devices, such as magnetic micro-electro-mechanical systems (MEMS), magnetic logics, spin electronics, or neuromorphic computing, *i.e.*, envisaging ultra-low power emulation of the biological synapse. Boosting the induced changes in magnetization, magneto-ionic motion and cyclability (endurance) continue to be key challenges to turn magneto-ionic phenomena into real applications. Here, we demonstrate that, without degrading cyclability, room temperature magneto-ionic motion in electrolyte-gated paramagnetic and fairly thick (> 100 nm) $Co_3O_4$ films largely depends on the configuration used to apply the electric field. In particular, magneto-ionic effects are significantly increased both in terms of generated magnetization (6 times larger: from **118.5** to **699.2 emu cm$^{-3}$**) and speed (35 times faster: from **33.1** to **1170.8 emu cm$^{-3}$ h$^{-1}$**) if the electric field is applied across a conducting buffer layer (grown underneath the $Co_3O_4$ films), instead of directly contacting $Co_3O_4$. This is attributed to a greater uniformity and strength of the applied electric field when using the conducting layer. These results may trigger the use of oxygen magneto-ionics into promising new technologies, such as magnetic MEMS or brain-inspired computing, which require endurance and moderate speeds of operation.

Current computers rely on Von Neumann's structural design in which the central processing unit and memory constitute different sub-devices bridged by the communication bus. This is not only detrimental to data processing speed but also to energy efficiency, making the search for new computing architectures central for future information technologies[1–6]. Neuromorphic computing relies on the use of devices that emulate the electrical behavior of the biological synapse, which is the memory and learning element of the brain, and has emerged as an alternative, which could render low-power information storage/processing[2–4,6]. Intense research is currently being carried out in the use of electronic-[2,3] and spintronic-based[4,5] approaches to mimic synapse's activity. However, these strategies are ultimately based on the use of electric currents (to generate magnetic fields through electromagnetic induction or spin-polarized electric currents via spin-transfer torque), involving a significant energy loss by heat dissipation through Joule effect[7,8]. Voltage control of magnetism (VCM), wherein magnetism is controlled via an applied electric field in place of an electric current, could potentially represent a significant breakthrough, envisaging ultra-low power emulation of the biological synapse[5]. Besides its potential use in memories and computation, magnetoelectric materials have also shown potential, in recent years, to be used in magnetic micro-electro-mechanical systems, such as biomagnetic sensors[9] or nanofluidics[10], where effects induced at moderate rates (*e.g.*, 1 s) are appealing to boost energy efficiency.

VCM has conventionally branched into three broad approaches, including (i) single-phase, multiferroic materials[11], (ii) inverse magnetostriction effects in piezoelectric/ferromagnetic heterostructures[12,13], and (iii) surface charge accumulation in ferromagnetic metals (*i.e.*, direct electric field effect)[14,15]. Each of the aforementioned VCM mechanisms has some drawbacks: multiferroics are limited in number, particularly at room temperature; strain-mediated heterostructures suffer from mechanical fatigue; and, in metals and alloys, electronic charging is only observed in ultra-thin ferromagnetic films.

Electric field-induced oxygen motion in magnetic materials (magneto-ionics) has recently revolutionized VCM since this mechanism may allow for a voltage-driven modulation of magnetic properties, such as coercivity, exchange bias field or magnetic anisotropy, to a level never reached by any other magnetoelectric means (*i.e.*, approaches (i), (ii) and (iii) above)[6,16–28]. Magneto-ionics, with magnetoelectric coupling efficiencies of the order of $10^3$ fJ/(V·m), deals so far with the lowest writing/retrieving energies (~$10^{-3}$ fJ/bit = 1 aJ/bit) to write/delete a bit (*i.e.*, information unit)[16,22]. This represents energies two and five orders of magnitude lower than those required in complementary metal oxide semiconductor (CMOS) technology (~$10^{-1}$ fJ/bit) and magnetic-based devices like magneto-resistive random access memories or hard disk drives (~$10^2$ fJ/bit), respectively[22].

Typically, magneto-ionic systems consist of layered heterostructures in a condenser-like configuration to apply voltage in solid state. Ferromagnetic metals, such as Co[16] or Fe[22], are the target materials grown adjacent to GdO$_x$ or HfO$_2$ layers, which act as ion reservoirs and, thereby, accepting or donating oxygen ions depending on the voltage polarity. Room temperature ionic motion is slow, involving times between $10^2$-$10^3$ s ($10^{-2}$-$10^{-3}$ Hz in frequency rate) to switch the magnetic state, such as the magnetic anisotropy easy axis from out-of-plane to in-plane and *vice versa* in ultra-thin Co layers by voltage-driven oxygen migration from a $GdO_x$ reservoir[16]. Therefore, alongside the applied voltage, these solid electrolytes usually require of high temperatures since ion migration is a thermally activated process[16–18,23,24]. In these magneto-ionic systems, the pristine ferromagnetic layer suffers from pronounced structural and compositional changes, leading to irreversibility[17] and, thus, poor cyclability[16]. Recently, via a proton-based approach, excellent endurance and $10^{-1}$ s (10 Hz) room temperature operation has been shown feasible in spite of certain instability since hydrogen retention is limited[23]. An alternative approach is the use of structural oxygen (self-contained in the magnetic material of interest), hence avoiding the need of external oxygen sources[20]. This has been shown in electrolyte-gated paramagnetic $Co_3O_4$ films, in which room temperature voltage-controlled on-off ferromagnetism has been achieved by electric switching of the oxidation state of cobalt (*i.e.*, voltage-driven reduction/oxidation), taking advantage of the defect-assisted voltage-driven migration of structural oxygen[20]. Even though this route still yields slow room temperature magneto-ionic motion, it shows outstanding stability and promising cyclability since the target is already oxidized.

Herein, by applying electric field using a conducting rather than an insulating buffer layer underneath the $Co_3O_4$ films (like previously reported results)[20], we demonstrate that, without degrading cyclability, room temperature magneto-ionic motion in electrolyte-gated, paramagnetic and fairly thick $Co_3O_4$ films (thicknesses above 100 nm) can be enhanced in terms of both generated magnetization (6 times larger) and speed (35 times faster). This is ascribed to a greater uniformity and strength of the applied electric field achieved when using the conducting buffer layer as working electrode instead of directly contacting the $Co_3O_4$ film. Our results, showing the importance of properly optimizing device design to apply electric field, could promote the use of oxygen magneto-ionics for brain-inspired computing[20,29] and new types of MEMS devices, which demand endurance and moderate speed of operation[30].

**Results**

Figs. 1a and 1b show the two types of film structures ($Co_3O_4$ (130 nm)/$SiO_2$ (20 nm)/[100]-oriented Si substrate and $Co_3O_4$ (130 nm)/TiN (170 nm)/[100]-oriented Si substrate, respectively) investigated in this work, aimed at unraveling the role of electric properties of the buffer layer underneath (*i.e.*, insulating for $SiO_2$ and conducting for TiN[31]: insulating *vs.* conducting configuration) in the magneto-ionic response of $Co_3O_4$.

Electrolyte-gating is used to apply voltage while performing in-plane vibrating sample magnetometry (VSM) –*i.e.*, magnetoelectric measurements–. A Pt wire is used as counter electrode, whereas the $Co_3O_4$ itself and the TiN are used as working electrodes for the $Co_3O_4$ (130 nm)/$SiO_2$ (20 nm)/substrate and $Co_3O_4$ (130 nm)/TiN (170 nm)/substrate sample configurations, respectively (see Methods for further details). The as-prepared $Co_3O_4$ samples (under no applied voltage) show virtually no ferromagnetic behavior in either the insulating (Fig. 1c) or the conducting configuration (Fig. 1d), in agreement with their paramagnetic nature at room temperature[20]. To investigate magneto-ionics in each configuration, the samples were subjected to –50 V for several hours and magnetic hysteresis loops of 25 min of duration were continuously recorded. After subjecting each sample to –50 V for 25 min (*i.e.*, upon the first hysteresis loop is recorded), the measurements show a clear hysteretic behavior, evidencing the emergence of ferromagnetism. The conducting configuration shows a remarkable increase in magnetization upon sweeping the first quadrant of the first hysteresis cycle, which doubles once the measurement reaches the fourth quadrant of the first loop. A much more gradual increase of the magnetization is observed in the insulating configuration. Fig. 1e shows the saturation magnetization ($M_S$) as a function of time (see Supplementary Figure 1 for further details on $M_S$ quantification). The magnetic moment scales monotonically with time for each configuration, but with a 6-fold larger increase between $SiO_2$ and TiN in the total magnetization (**118.5** to **699.2 emu cm$^{-3}$**, respectively) reached after magneto-ionic motion has stabilized. Furthermore, the time scale for ferromagnetism generation ("on" state) in the conducting configuration is significantly faster than in the insulating configuration. To compare properly the rate of "on" switching, this magnetization increase is determined by a linear fit of the $M_S$ vs. $t$ plot evaluated during the first 36 min of voltage application (wherein $M_S$ in the TiN configuration fully saturates). The rates are **33.1** and **1170.8 emu cm$^{-3}$ h$^{-1}$**, showing that the use of a conducting buffer layer enhances ion migration by a factor 35 with respect to the insulating buffer layer.

Looking at the *M* (magnetization)-*H* (applied magnetic field) loops (Figs. 1c and 1d), there are also marked shape differences. The conducting configuration exhibits more square-shaped and more tilted cycles (*i.e.*, it has a more "easy axis" character) than the insulating configuration. To examine the shape of the *M-H* loops, the squareness, defined as the ratio between the remnant magnetization ($M_R$) and $M_S$ ($M_R/M_S$), and the slope of the hysteresis loop at the coercive field ($H_C$) normalized to $M_S$ (d$M$/d$H$ ($H = H_C$) $M_S^{-1}$) have been calculated for both the descending and ascending branches of the measured hysteresis loops (Fig. 1f). The conducting configuration exhibits higher $M_R/M_S$ ratios and slopes at $H_C$ throughout the time the voltage was applied, in concordance with more square-shaped and more tilted loops (*i.e.*, narrower distribution of coercive fields)[32].

To further examine the nature of the electric field experienced by the $Co_3O_4$ samples, COMSOL simulations were performed to model the initial voltage distributions for each configuration upon electrolyte-gating (see Methods for further simulation details). In Figs. 1g and 1h, electric contact to the working electrodes ($Co_3O_4$ and TiN for the insulating and conducting configurations, respectively) is made at the top of the left plane which represents the samples, whereas the right plane corresponds to the counter electrode (*i.e.*, Pt wire).

Clear differences can be seen in the equipotential lines for the insulating (Fig. 1g) and conducting (Fig. 1h) configurations. In the insulating configuration, the dielectric nature of SiO$_2$ and limited electric conductivity of Co$_3$O$_4$[33] manifest in a non-homogeneous voltage distribution along the vertical extent of the Co$_3$O$_4$ film, showing a weaker and less uniform applied electric field as the distance from the electric contact is increased. Conversely, in the other configuration, the conducing nature of TiN results in a nearly uniform voltage distribution along the vertical cross-section of the sample, which gives rise to a larger and better defined electric field along the direction perpendicular to the Co$_3$O$_4$ film plane. In contrast to the dielectric configuration, the whole Co$_3$O$_4$ film is activated for magneto-ionic motion (see Supplementary Figure 2).

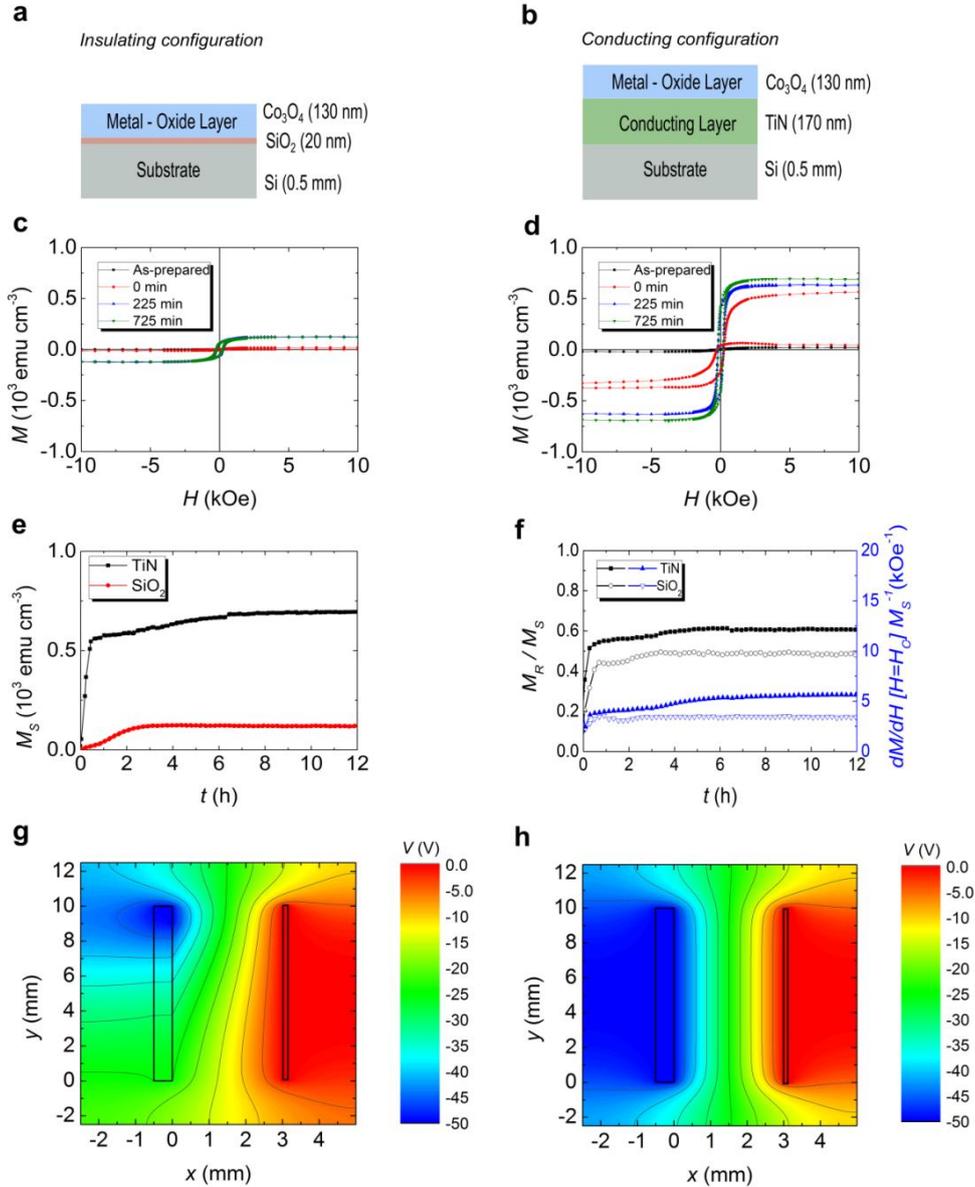

**Fig. 1** Sample configurations: (**a**) Co$_3$O$_4$ on SiO$_2$ (insulating configuration) and (**b**) Co$_3$O$_4$ on TiN (conducting configuration). (**c**) and (**d**) consecutive hysteresis loops under −50 V gating for the insulating and conducting configurations, respectively, taken by in-plane vibrating sample magnetometry. (**e**) Time evolution of the saturation magnetization ($M_S$ vs. $t$) and (**f**) squareness ($M_R/M_S$) & slope of hysteresis loop at $H_C$ normalized to $M_S$ for each configuration. (**g**) and (**h**) show the COMSOL simulations of the initial voltage distribution at the moment in which the insulating and the conducting configurations, respectively, are electrolyte-gated (equipotential lines are drawn).

To assess the degree of structural and compositional change that $Co_3O_4$ undergoes with voltage for the two investigated configurations, cross-section lamellae of the pristine and treated $Co_3O_4$ films were prepared and characterized by high-angle annular dark-field scanning transmission electron microscopy (HAADF-STEM) and electron energy loss spectroscopy (EELS), respectively (Fig. 2).

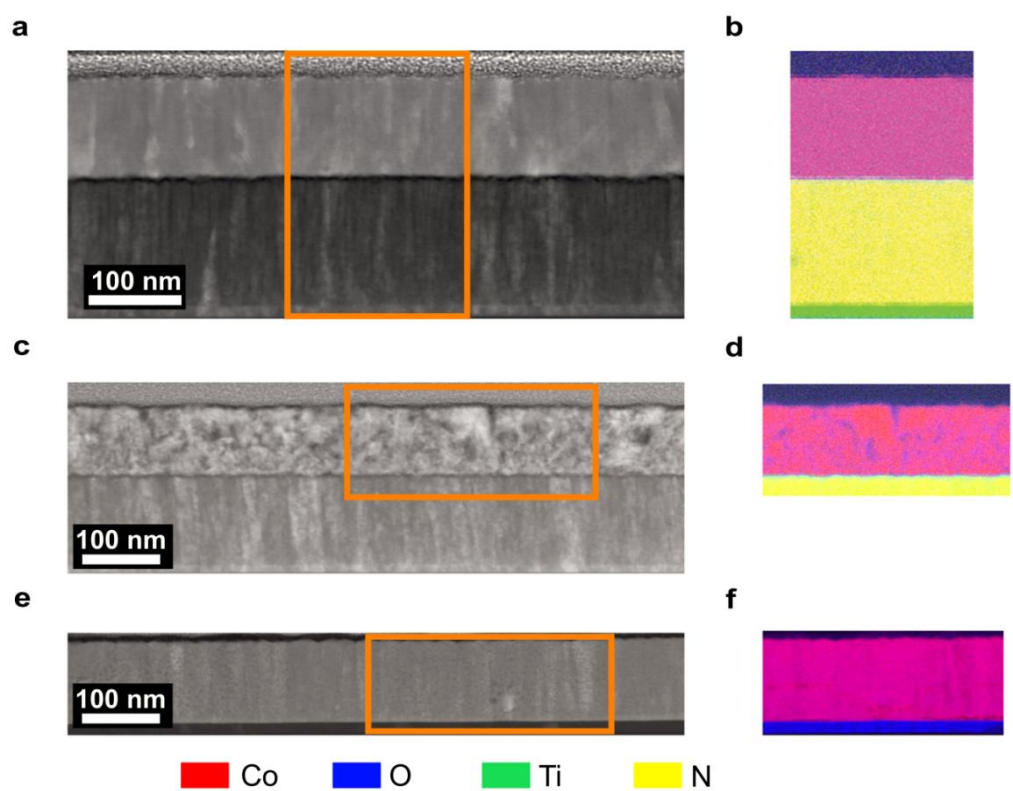

**Fig. 2** Structural and compositional characterization by high-angle annular dark-field scanning transmission electron microscopy (HAADF-STEM) and electron energy loss spectroscopy (EELS), respectively. (**a**–**b**), (**c**–**d**), and (**e**–**f**) are HAADF-STEM images and elemental EELS mappings of the areas marked in orange corresponding to pristine $Co_3O_4$/TiN, $Co_3O_4$/TiN negatively biased at −50 V for 80 min, and $Co_3O_4$/SiO$_2$ also negatively biased −50 V for 80 min, respectively. Colors corresponding to each element for the EELS analyses are depicted at the bottom of the figure.

The morphology of the pristine sample grown on TiN (Fig. 2a) shows regular, columnar-shaped grains as it happens in $Co_3O_4$ deposited by atomic layer deposition on $SiO_2$[20]. This morphology remains rather unaltered after treating the $Co_3O_4$ film deposited on SiO$_2$ with −50 V for 80 min (Fig. 2e). On the contrary, the $Co_3O_4$ morphology in the conducting configuration treated at −50 V for 80 min shows no columnar grains consistent with a more nanostructured $Co_3O_4$ phase (Fig. 2c and Supplementary Figure 3).

To locally quantify the Co/O distribution, Co and O EELS mappings were conducted for the as-grown films and the samples treated at −50 V for 80 min for both insulating and conducting configurations (Fig. 2). Co (red) and O (blue) are homogeneously distributed in the as-grown sample with conducting configuration (Fig. 2b) and nearly homogeneously distributed in the treated sample with the insulating configuration (Fig. 2f), which sharply contrasts with the sample treated under −50 V in the conducting configuration (Fig. 2d). The corresponding Co (red) and O (blue) EELS mappings reveal the presence of Co-rich and O-rich areas due to voltage-driven ion migration. In contrast to the sample grown on $SiO_2$, electrolyte-gating of the $Co_3O_4$ sample grown on TiN results in $O_2$ bubbling, evidencing that, on top of oxygen redistribution within the film[20], oxygen is also released into the liquid medium, which acts as an oxygen sink.

Further structural characterization was carried out by $\theta/2\theta$ X-ray diffraction (XRD), high resolution transmission electron microscopy (HRTEM) and variable energy positron annihilation spectroscopy (VEPAS). The XRD patterns of the as-prepared samples are consistent with a textured $Co_3O_4$ phase along (1 1 1), (2 2 2) and (3 1 1) planes. Upon electrolyte-gating at −50 V for 80 min, the intensity of the (1 1 1) and (2 2 2) planes strongly decreases while that of the (3 1 1) planes reduces only slightly. For the $Co_3O_4$ sample grown on TiN, the peak corresponding to (1 1 1) planes fully vanishes after the application of this negative voltage (Fig. 3a). Furthermore, as seen in the detailed XRD view of Fig. 3b, the conducting configuration shows the emergence of a new peak after gating at −50 V for 80 min, which is consistent with the diffraction from (0 0 2) planes of hexagonal close-packed Co (HCP-Co).

Moreover, high resolution transmission electron microscopy (HRTEM) was performed in the cross-section of a $Co_3O_4$ film grown on TiN and treated at −50 V (Fig. 3c). The inset shows the fast Fourier transform of the area marked with a red rectangle, which results in three well-defined spots highlighted in red circles and numbered 1, 2 and 3. The corresponding interplanar distances are 1.991, 1.920 and 2.535 Å, respectively. The interplanar distance of 1.920 Å is unambiguously ascribed to (1 0 1) HCP-Co (ICDD JCPDF 00-005-0727), whereas 1.991 Å could be associated with either (0 0 2) HCP-Co or (4 0 0) $Co_3O_4$ (ICDD JCPDF 00-009-0418). However, since the as-prepared $Co_3O_4$ films do not exhibit traces of (4 0 0) planes and (0 0 2) HCP-Co is observed by XRD, the interplanar distance of 1.991 Å is likely to belong to HCP-Co. Finally, 2.535 Å is consistent with an O-deficient (3 1 1) $Co_3O_4$ (ICDD JCPDF 00-009-0418) phase[20].

To examine the microstructure of the as-prepared films at atomic level, variable energy positron annihilation spectroscopy (VEPAS) was performed (Fig. 3d). Both low and high electron momentum fraction ($S$ and $W$, respectively) as a function of positron implantation energy, $E_p$, virtually overlap up to the first 50 nm in depth. The differences at further depths of the film are essentially due to the different chemical nature of the buffer layer. This indicates that the as-prepared films grown on both substrates have similar amount and type of defects, independently of the substrate they are grown on.

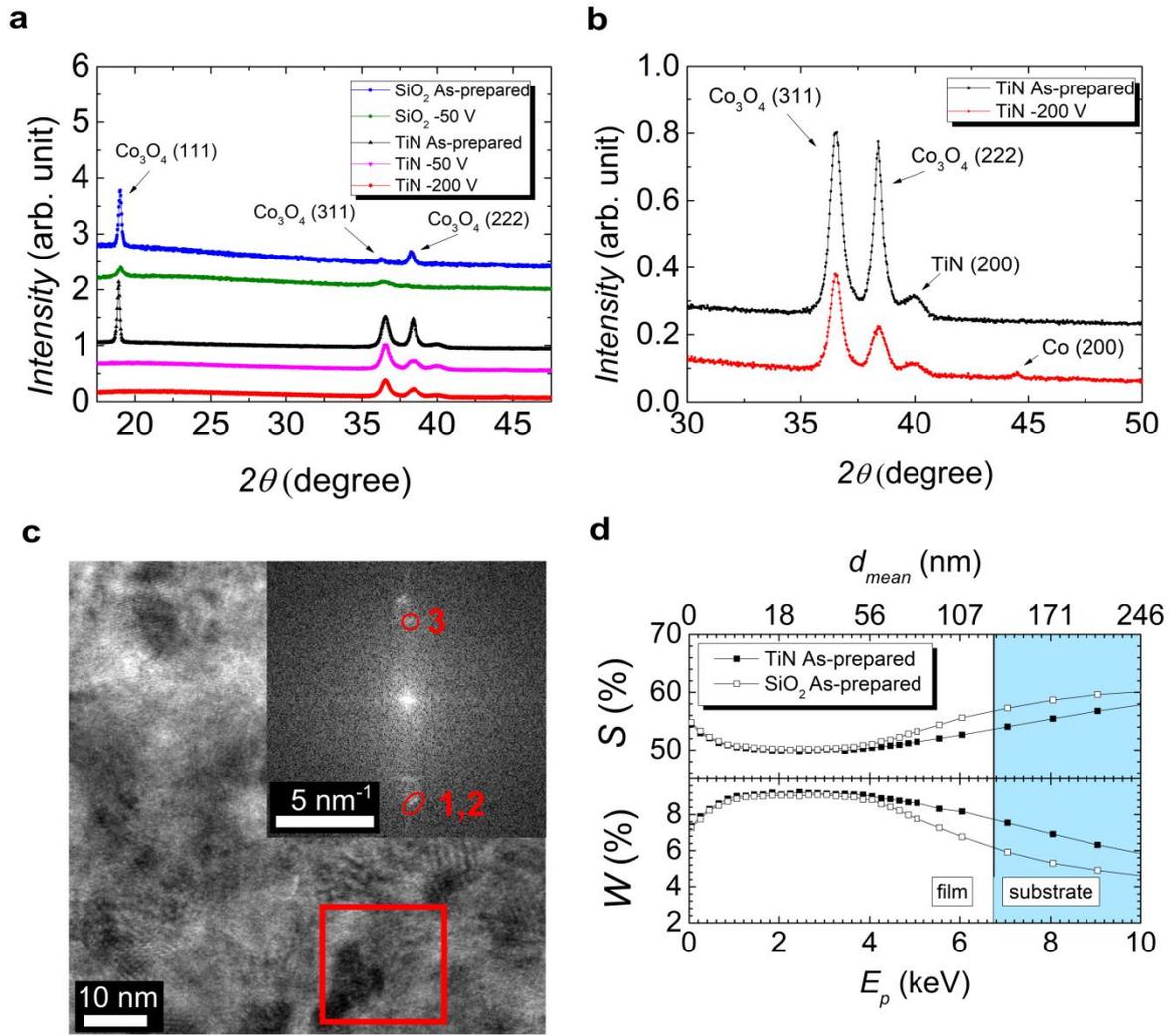

**Fig. 3** Structural characterization by $\theta/2\theta$ X-ray diffraction (XRD), high resolution transmission electron microscopy (HRTEM) and variable energy positron annihilation spectroscopy (VEPAS). (**a**) $\theta/2\theta$ XRD diffraction patterns of the as-prepared and treated samples at −50 V, and at −200 V for the $Co_3O_4$ grown on TiN. (**b**) Detailed view of the XRD patterns corresponding to the $Co_3O_4$ grown on TiN. (**c**) HRTEM image of the cross-section of a $Co_3O_4$ film grown on TiN and treated at −50 V. The inset shows the fast Fourier transform of the area marked with a red rectangle. (**d**) Low and high electron momentum fraction (*S* and *W*, respectively) as a function of positron implantation energy, $E_p$, for the as-prepared samples. "$d_{mean}$" stands for average depth.

The onset voltage for magneto-ionic motion and cyclability has also been investigated for both configurations (Fig. 4). To determine the onset voltage, the gating was monotonically decreased in steps of −2 V to observe when the system started to display ferromagnetic behavior. Afterwards, the voltage polarity was reversed to test the cyclability of the magneto-ionic effect. The $Co_3O_4$ film grown on TiN exhibits an onset voltage of −4 V and requires of +50 V to fully recover the pristine paramagnetic state. Conversely, the insulating configuration shows an onset voltage of −10 V and requires of +10 V to recover the initial state, in agreement with previously reported results on the same configuration but with a thicker $SiO_2$ buffer layer[20].

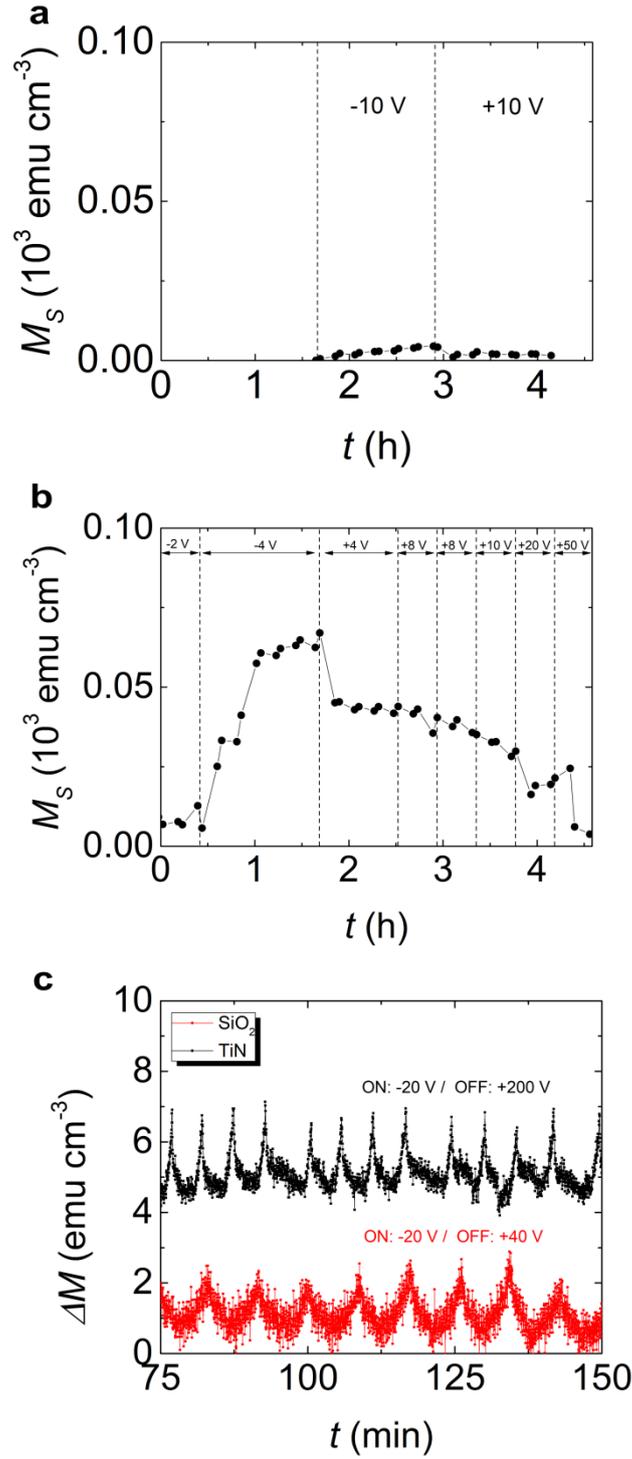

**Fig. 4** (**a**) and (**b**) onset/recovery behavior of the insulating and conducting configurations, respectively. (**c**) Cyclability for both sample configurations (−20 V/+40 V and −20 V/+200 V pulses for the insulating and conducting configurations, respectively).

While the onset/recovery process is repeatable, to perform cyclability tests the applied biases were increased in order to enhance the magneto-ionic signal-to-noise ratio and, thus, better observe endurance. 30 cycles were performed using voltage pulses of −20 V/+40 V and −20 V/+200 V for the insulating and conducting configurations, respectively (Fig. 4c). The change in magnetization relative to the background magnetization is observed to be repeatable in time scale and magnetization quality, suggesting suitable reproducibility for long term use.

**Discussion**

The role of the electric properties of the buffer layer (*i.e.*, insulating $SiO_2$ or conducting TiN) in the magneto-ionic behavior of $Co_3O_4$ thin films has been investigated. As seen in Fig. 1, upon electrolyte-gating at −50 V, the use of a conducting rather than an insulating buffer layer boosts magneto-ionics in terms of both long-term generated magnetization (6-fold relative increase: from **118.5** to **699.2 emu cm$^{-3}$**) and initial magneto-ionic motion (35 times faster: from **33.1** and **1170.8 emu cm$^{-3}$ h$^{-1}$**).

These different behaviors are already revealed by the *M-H* hysteresis loops which show pronounced differences in shape. The conducting configuration results in more square-shaped loops with larger squareness and more tilted branches (*i.e.*, narrower distribution of coercive fields and, thus, higher slopes at $H_C$), evidencing a more "easy-axis" nature than the insulating configuration. This is consistent with the generation of more uniform ferromagnetic regions in the $Co_3O_4$ film (with better defined shape anisotropy) when a conducting buffer layer is used, in agreement with the COMSOL simulations of Figs. 1g and 1h. In the insulating configuration, the interplay between the dielectric nature of $SiO_2$ and the limited electric conductivity of $Co_3O_4$ manifests in a non-homogeneous voltage distribution along the $Co_3O_4$ film, showing a weaker and less uniform applied electric field as the distance from the electric contact increases. Conversely, in the other configuration, the conducing nature of TiN results in a nearly uniform voltage distribution along the vertical cross-section of the sample, which gives rise to a larger and better defined electric field along the direction perpendicular to the $Co_3O_4$ film plane. In contrast to the dielectric configuration, the $Co_3O_4$ film on TiN is fully and homogeneously activated alongside its complete vertical extent. This results in a well-defined path with a nearly full perpendicular electric field component for magneto-ionic motion in the conducting configuration. On the contrary, in the insulating configuration, due to the presence of a vertical component caused by the limited electric conductivity of $Co_3O_4$, the strength and speed of magneto-ionic motion is hindered. This is in concordance with a broader size distribution of ferromagnetic regions in the electrolyte-gated $Co_3O_4$ film on $SiO_2$, in agreement with the magnetometry results which indicate lower squareness values and less tilted branches (*i.e.*, broader distribution of coercive fields). This is also evidenced by the evolution of coercivity with time for the consecutive loops taken while electrolyte-gating both configurations at −50 V (Supplementary Figure 4). Whereas the insulating configuration results in a monotonic increase of $H_C$ with time, the conducting configuration shows a maximum at the very beginning. This maximal behavior resembles the typical dependence of coercivity with particle size in magnetic systems, consistent with a scenario in which a more homogeneous generation of ferromagnetic regions occurs, uniformly evolving in size, likely starting from a superparamagnetic behavior, followed by a single domain state (maximum of $H_C$) and ending with a multi-domain configuration[32].

The effect of configuration on the compositional and structural properties of $Co_3O_4$ is clearly observed in Fig. 2. The morphology of the pristine samples shows regular, columnar-shaped grains (Fig. 2a and Supplementary Figure 3) and homogeneous composition (Fig. 2b). This largely remains upon treating the $Co_3O_4$ film deposited on $SiO_2$ with −50 V for 80 min (Figs. 2e and 2f). Conversely, the morphology of the conducting configuration treated at −50 V for 80 min shows almost no columnar grains and a highly nanostructured $Co_3O_4$ phase (Fig. 2c and Supplementary Figure 3) with Co and O segregation (Fig. 2d). This indicates that the conducting configuration can electrically modulate ion migration at much higher strengths. This is further confirmed by XRD and HRTEM which show traces of metallic Co only for the conducting configuration (Fig. 3). Even though the as-prepared $Co_3O_4$ films show similar crystallographic features regardless buffer layer, further structural characterization to examine the local microstructure of the as-prepared samples was carried out by VEPAS. As can be seen in Fig. 3d, both low and high electron momentum fraction ($S$ and $W$, respectively) as a function of positron implantation energy, $E_p$, virtually overlap, indicating that $Co_3O_4$ either grown on $SiO_2$ or TiN exhibits analogous defect environment, ruling out minor microstructure differences in $Co_3O_4$ as the origin of the observed magneto-ionic effects. Nevertheless, voltage-driven ion diffusion seems to be facilitated along (1 1 1) planes as evidenced by the strong decrease in intensity of the (1 1 1) and (2 2 2) planes, with the (1 1 1) planes fully vanishing for the $Co_3O_4$ sample grown on TiN (Fig. 3a).

As seen in Fig. 4, the minimum voltage bias required to perform an onset/recovery cycle is asymmetric for the $Co_3O_4$ film grown on TiN (−4 V/+50 V), while it is symmetric for the $Co_3O_4$ film grown on $SiO_2$ (−10 V/+10 V). The onset bias is significantly larger for the conducting configuration (−4 V (conducting TiN) vs. −10 V (insulating $SiO_2$)). In contrast to the insulating configuration, the use of a conducting buffer layer activates the whole $Co_3O_4$ sample, resulting in a more intense and better defined perpendicular electric field and, thus, in enhanced magneto-ionic effect and motion. $O_2$ bubbling is noticeably observable only in the conducting configuration, evidencing that, on top of oxygen redistribution[20], oxygen is released into the liquid electrolyte, which acts as an oxygen sink, and as an oxygen reservoir due to the oxygen solubility in propylene carbonate. The voltage asymmetry in the conducting configuration can be linked to $O_2$ bubbling since propylene carbonate reaches O supersaturation and the oxygen forming bubbles cannot be recovered.

Cycling (Fig. 4) further corroborates the faster magneto-ionic rates of the conducing configuration, particularly during the generation of the "on" states. Time span in the insulating configuration has been enlarged to reach a suitable signal-to-noise ratio, as a consequence of the slower magneto-ionic kinetics. For the conducting configuration, cyclability is lost when lower voltages (higher in absolute value but negatively biased) are applied (e.g., −50 V) due to strong irreversible $O_2$ bubbling.

In summary, we have investigated the role of the electric properties of the substrate in the magneto-ionic behavior of $Co_3O_4$ thin films. Polycrystalline 130 nm-thick $Co_3O_4$ films have been grown by atomic layer deposition on either insulating $SiO_2$ or conducting TiN buffer layers. The use of a conducting rather than contacting from the top $Co_3O_4$ (when deposited onto an insulating substrate) boosts magneto-ionics in terms of both generated magnetization (6-fold increase: from **118.5** ($Co_3O_4/SiO_2$) to **699.2 emu cm$^{-3}$** ($Co_3O_4$/TiN)) and magneto-ionic rates (35 times faster: from **33.1** ($Co_3O_4/SiO_2$) to **1170.8 emu cm$^{-3}$ h$^{-1}$** ($Co_3O_4$/TiN)). Upon gating, transmission electron microscopy and electron energy loss spectroscopy show the emergence of Co-rich areas at a greater intensity for the $Co_3O_4$ grown on a conducting substrate. Magnetization measurements also show a marked increase in the squareness ratio and a decrease in the switching field distribution of the hysteresis loops from $Co_3O_4$ deposited on the conducting layer, evidencing the generation of more uniform ferromagnetic regions. This dissimilar behavior between the use of either an insulating or a conducting substrate arises from the intensity and uniformity of the electric field, which are maximized when using a conducting substrate while preserving stability and endurance. These results demonstrate the importance of the specific device design (and, in particular, the use of conducting substrates to apply electric field in electrolyte-gated oxide materials) in order to optimize the strength and speed of the magneto-ionic effect. Our results prompt the way to make oxygen magneto-ionics feasible for practical applications in fields like neuromorphic and stochastic computing or magnetic MEMS, where high frequencies are not necessarily required for device engineering[29].

**Methods**

**Sample preparation**

130 nm thick $Co_3O_4$ films were grown on two substrates: i) thermally oxidized non-doped Si wafers ($SiO_2$ (20 nm)/[1 0 0]-oriented Si (0.5 mm)), and ii) non-doped Si wafers coated with a TiN buffer layer (TiN (170 nm)/[1 0 0]-oriented Si (0.5 mm)). Deposition was carried out by plasma enhanced atomic layer deposition as described in references 20, 33 and 34.

**Magnetoelectric characterization**

Magnetic measurements under electrolyte gating (*i.e.*, magnetoelectric characterization) were carried out at room temperature in a vibrating sample magnetometer from Micro Sense (LOT-Quantum Design), with a maximum applied magnetic field of 2 T. The sample was mounted in a homemade electrolytic cell filled with anhydrous propylene carbonate with $Na^+$ solvated species (5 - 25 ppm), and the magnetic properties were measured along the film plane after applying different voltages, using an external Agilent B2902A power supply, between the sample and the counter-electrode in a similar fashion of that presented in references 15, 20 and 28. The $Na^+$ solvated species in the electrolyte are aimed at reacting with any traces of water[15]. The magnetic signal was normalized to the area of the sample exposed to the electrolyte during the voltage application process. All hysteresis loops were background-corrected and the correction was carried out at high fields (*i.e.*, fields always far above saturation fields) to eliminate linear contributions (paramagnetic and diamagnetic signals).

**Structural and compositional measurements**

$\theta/2\theta$ X-ray diffraction (XRD) patterns were recorded on a Philips X'Pert Powder diffractometer with a PIXcel[1D] detector using Cu $K_\alpha$ radiation.

High resolution transmission electron microscopy (HRTEM), high-angle annular dark-field scanning transmission electron microscopy (HAADF-STEM) and electron energy loss spectroscopy (EELS) were performed on a TECNAI F20 HRTEM /STEM microscope operated at 200 kV. Cross sectional lamellae were prepared by focused ion beam and placed onto a Cu transmission electron microscopy grid.

Variable energy positron annihilation spectroscopy (VEPAS)[35,36] was used to investigate depth-resolved open volume defects at the Slow-Positron System of Rossendorf (SPONSOR) beamline, which provides monoenergetic but variable energy positron beam.

**Modelling**

A simulation of the charge distribution in each of the two systems was performed using COMSOL finite element analysis software. Estimation of the charge distribution considered charge conservation ($\nabla J = 0$), Ohm's law ($J = \sigma E$), and Gauss' law ($E = -\nabla V$). The geometry of the system was modeled in 2D to minimize computational needs. The geometry includes the silicon substrate (0.5 mm), the sample film ($Co_3O_4$ and buffer layer: either $SiO_2$ or TiN), contact layer (In solder onto Cu wire), a platinum counter electrode, all set in an electrochemical chamber filled with propylene carbonate. The sample film has been unified to enable a tetragonal adaptive mesh within the computer memory resources. The effective electric conductivity (*i.e.*, electric conductance) has been modeled by weighing the different sub-layer constituting within the sample stack and normalizing relative to the thicknesses of each layer in the model. In the insulating configuration the silicon substrate is considered coated with a SiO$_2$ layer, electrically isolated from the sample. The sample essentially consists of a 130 nm Co$_3$O$_4$ layer on top the substrate. A value of electrical conductivity of 20 S/m for Co$_3$O$_4$ has been measured via a 4-probe van der Pauw measurement, in agreement with values from the literature[33].

The electrical conductivity for TiN was taken from literature, 3×10$^5$ S/m[31]. In the conducting configuration, the 170 nm of TiN and the 130 nm of Co$_3$O$_4$ are taken in combination. The large electrical conductivity of the TiN dominates the effective electrical conductivity of the sample section. Simulations shown are calculated for the system under –50 V gating voltage.

**Data availability**

The data used in this paper are available from the corresponding authors upon request.

**Acknowledgements**

Financial support by the European Research Council (SPIN-PORICS 2014-Consolidator Grant, Agreement Nº 648454), the Spanish Government (MAT2017-86357-C3-1-R and associated FEDER) and the Generalitat de Catalunya (2017-SGR-292 and 2018-LLAV-00032) is acknowledged. This work was partially supported by the Impulse-und Net-working fund of the Helmholtz Association (FKZ VH-VI-442 Memriox), and the Helmholtz Energy Materials Characterization Platform (03ET7015).

**Author contributions**

**E.M.** had the original idea and led the investigation. **J.d.R.**, **J.S.** and **E.M.** designed the experiments. **L.H.**, **J.D.** and **C.D.** synthesized the materials. **A.Q.** designed the sample holder to carry out magnetometry under voltage application (*i.e.*, magnetoelectric measurements). **J.d.R.**, **J.S.** and **E.M.** carried out the magnetoelectric measurements and analyzed the data.

**J.d.R.**, **A.Q.** and **E.M.** performed the XRD and TEM characterization and analyzed the corresponding data. **A.L.** and **L.A.** conducted the COMSOL simulations. **J.S.**, **B.M.** and **J.L.C.** performed the electrical characterization of the samples. **M.O.L.**, **M.B.** and **A.W.** characterized the samples by VEPAS and analyzed the data. All authors discussed the results and commented on the article. The article was written by **J.d.R.**, **J.S.** and **E.M.**

**Competing financial interests**

The authors declare no competing financial interests.

**Supplementary Information**

**Supplementary Figure 1**

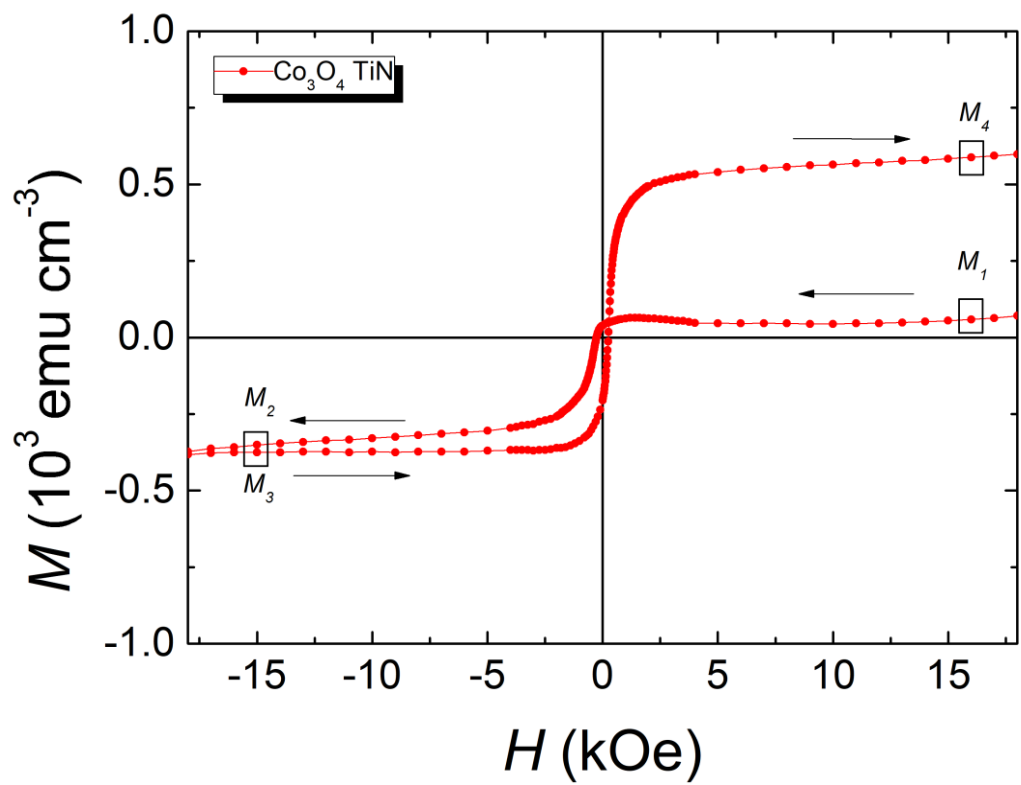

Saturation magnetization ($M_S$) from hysteresis loops. The cartoon is based on the first hysteresis loop of the $Co_3O_4$ film grown on TiN (conducting configuration) upon electrolyte-gating at −50 V. Since sample magnetization is continuously evolving, the $M_S$ is taken at ±15 kOe for each of the quadrants of the hysteresis loop (*i.e.*, $M_1$, $M_2$, $M_3$ and $M_4$).

**Supplementary Figure 2**

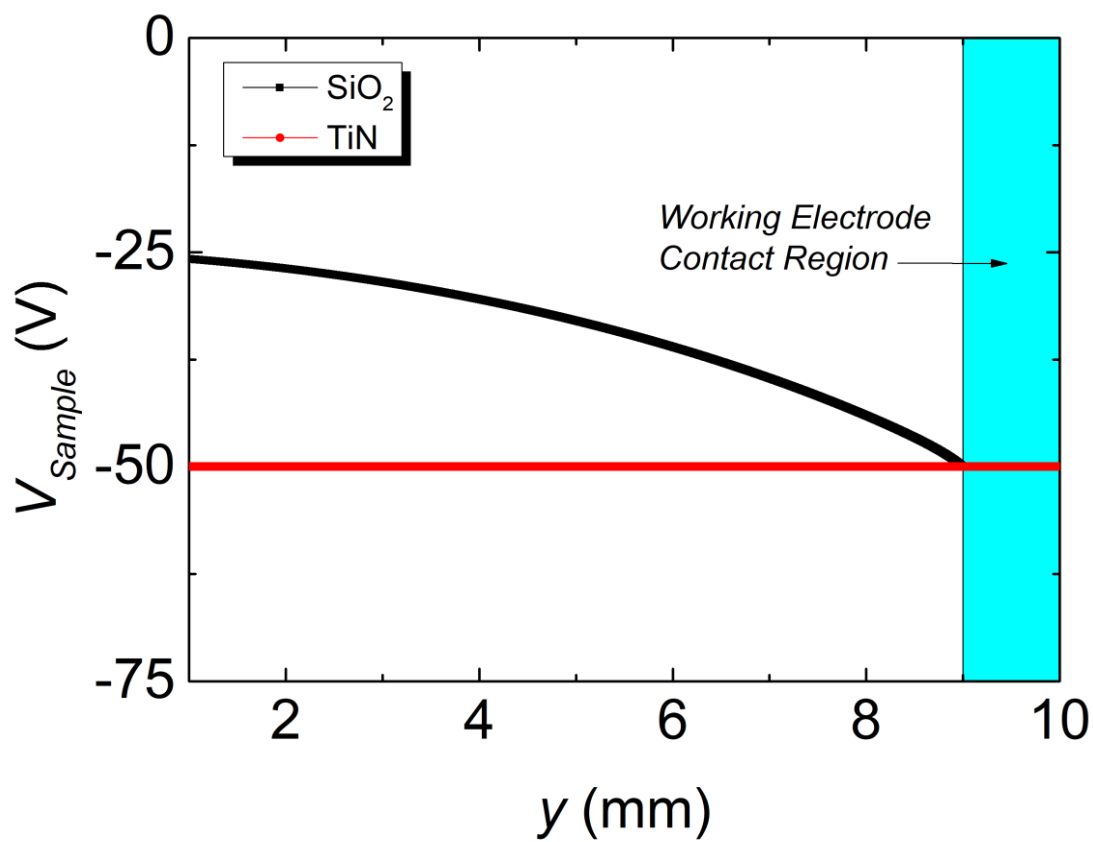

Electric potential (upon electrolyte-gating at −50 V) at the bottom of the $Co_3O_4$ layer along the length of the sample for both configurations: $Co_3O_4$ on either $SiO_2$ (insulating) or TiN (conducting).

**Supplementary Figure 3**

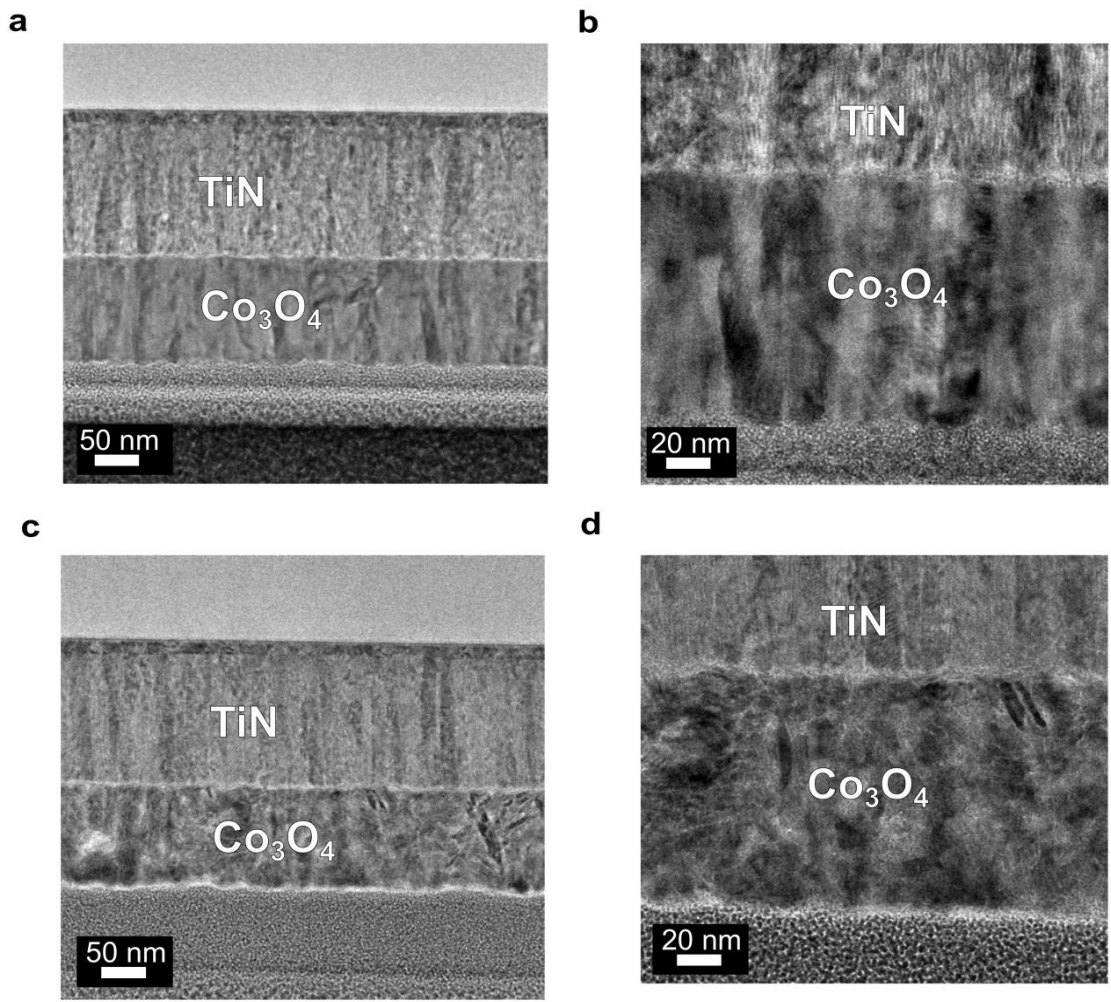

(**a**) and (**b**) are transmission electron microscopy (TEM) images of the cross-section of an as-prepared $Co_3O_4$ film grown on TiN, and (**c**) and (**d**) are TEM images corresponding to the same sample but treated at −50 V for 80 min.

**Supplementary Figure 4**

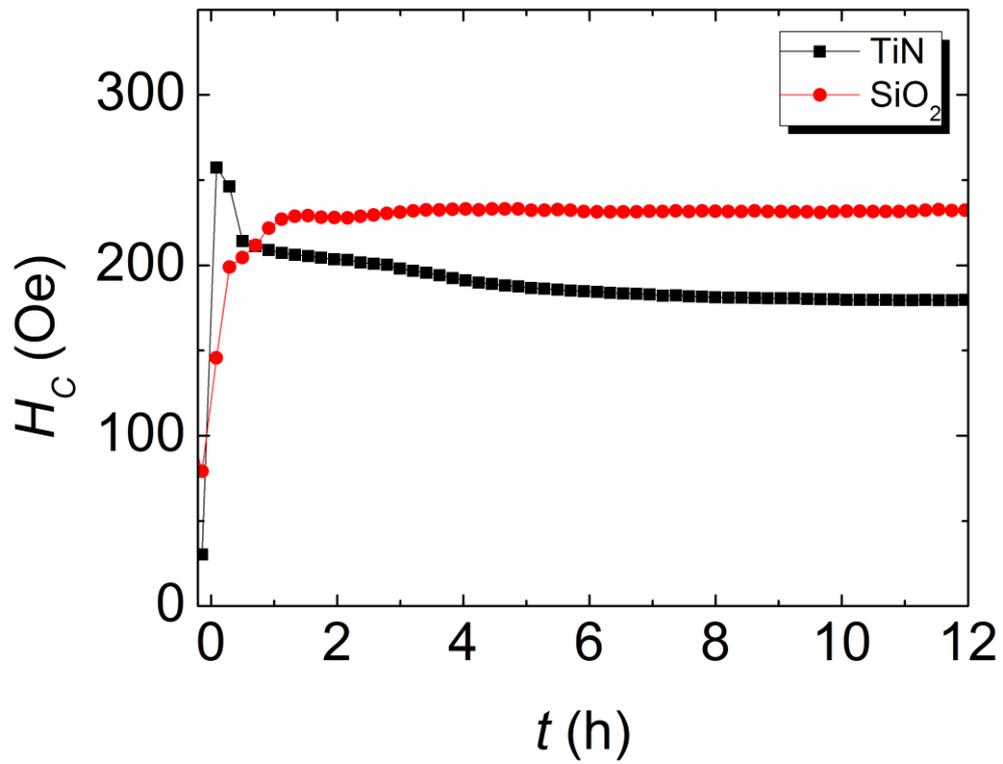

Coercive field ($H_C$) evolution with time for $Co_3O_4$ films grown on either $SiO_2$ (insulating configuration) or TiN (conducting configuration) subjected to a voltage of −50 V for several hours.